**Single crystal growth of MgB$_2$ and thermodynamics of Mg-B-N system at high pressure**

*J.Karpinski[1], S.M.Kazakov[1], J.Jun[1], M.Angst[1], R.Puzniak[2], A.Wisniewski[2], P.Bordet[3]

[1]*Solid State Physics Laboratory ETH, 8093-Zürich, Switzerland,*
[2]*Institute of Physics, Polish Academy of Sciences, Warsaw, Poland,*
[3]*Laboratoire de Cristallographie C.N.R.S. Grenoble, France*

**Abstract**

Single crystals of MgB$_2$ have been grown at high pressure via the peritectic decomposition of MgNB$_9$. The crystals are of a size up to 1.5x0.9x0.2mm$^3$ with a weight up to 230μg, and typically have transition temperatures between 37 and 39 K with a width of 0.3-0.5 K. Investigations of the *P-T* phase diagram prove that the MgB$_2$ phase is stable at least up to 2190$^o$C at high hydrostatic pressure in the presence of Mg vapor under high pressure. Small variations of *T$_c$* are caused by doping with metal elements from the precursor or annealing of defects during the crystal growth process.

**Introduction**

MgB$_2$ recently attracted a lot of interest due to the anisotropic two-band electronic structure, leading to the presence of two distinct energy gaps [1,2]. This causes unconventional properties, like the high critical temperature of 39 K and a temperature and field dependent anisotropy, which does not fit to the Ginsburg-Landau model with a constant anisotropy of the upper critical field, equal to the anisotropy of the penetration depth and the coherence length [3,4]. Anisotropic properties have to be studied on single crystals. Soon after the discovery of superconductivity in MgB$_2$, several laboratories started to work on single crystal growth. Unfortunately, conventional methods of crystal growth, like a growth from high temperature solutions in metals (Al, Mg, Cu etc.) at ambient pressure, used for other borides, did not work for MgB$_2$. Since MgB$_2$ melts non-congruently, it is also not possible to grow crystals from a stoichiometric melt. MgB$_2$ crystals can be grown from a solution in Mg at high pressure, or from the vapor phase. Two methods result in submilimeter size single crystals: high pressure growth using an anvil technique [4-7] and heating of a mixture of Mg and B in closed metal container [8-10]. Larger crystals have been obtained using the first method. The solubility of $MgB_2$ in Mg is extremely low at temperatures below the boiling temperature of Mg ($1107^oC$) at ambient pressure [Fig. 1a)], therefore crystals have to be grown at much higher temperature or by using another solvent. At higher temperatures at ambient pressure, Mg does not exist as a condensed phase, and therefore, in order to grow crystals from a solution in Mg, the pressure has to be increased. According to the calculated *P-T* phase diagram [11] [shown in Fig.1b)], crystal growth from a solution has to be performed in pressure and temperature conditions above the boiling line of Mg, which is the border between two regions in Fig.1b): Liquid+$MgB_2$ and Gas+$MgB_2$. However, for the stabilization the $MgB_2$ phase at temperatures above $1000^oC$, not only high partial pressure of Mg vapor is necessary, but also high hydrostatic pressure. There are several major problems, which have to be solved in order to grow crystals:

a) The reactivity of the crucible material: At $T>1000^oC$, molten Mg is very aggressive towards all materials and destroys the crucible after a few hours.

b) All metals used as a solvent form mixed compounds with Mg or $MgB_2$, which makes crystal growth of pure $MgB_2$ impossible from any solvent other than Mg.

c) The solubility of $MgB_2$ in Mg is very small at low temperatures and therefore high temperature are necessary. However, the partial pressure of Mg vapor above molten Mg increases with temperature and at $1500^oC$ is of the order of 50 bar.

d) $MgB_2$ decomposes above $1000^oC$, at ambient pressure.

In this paper, we will present the results of crystals growth experiments and the influence of crystal growth conditions and purity of the precursor on $T_c$. The high quality of the crystals allowed measurements of various physical properties, which are published in separated papers of this issue: magnetic studies of anisotropic properties and *H-T* phase diagram [3,4], scanning tunneling spectroscopy [12], point contact spectroscopy [13] and optical studies [14]. For optimization of crystal growth a knowledge of the thermodynamics of the Mg-B-N system is required. Here we will also present the *P-T* phase diagram of the Mg-B-N system, giving evidence of the stability of $MgB_2$ up to $2190^oC$ at high pressure, almost $1000^oC$ higher than published up to now.

**Crystal growth experiment**

Two methods of high-temperature solution crystal growth of $MgB_2$ have been investigated:

a) Crystal growth from a solution in Al, Mg, Cu, and mixtures of these metals, at high temperatures up to 1700°C at Ar gas pressure 1 kbar<$P_{Ar}$<14 kbar.

b) Crystal growth from a solution in Mg at high temperatures up 2190°C, using a cubic anvil technique with a solid pressure medium (10 kbar<$P$<35 kbar).

For experiments at Ar pressure, a mixture of Mg and B has been placed in a BN crucible in a high pressure furnace. The experiments have been performed at $P_{Ar}$=1 kbar, at temperatures from 1450 up to 1700°C. The experiment at $P_{Ar}$=14 kbar was done at 1250°C. As a result, hexagonal black crystals of the new nitride $MgNB_9$ were obtained, but no $MgB_2$ appeared. The source of nitrogen was the BN crucible. The growth of $MgB_2$ crystals at Ar pressure was also impossible from solutions containing Cu and Al, because Cu, used as a solvent, forms mixed compounds with Mg, while Al substitutes Mg in $MgB_2$.

Crystal growth experiments, using the anvil system, have been performed in the pressure range of 10-30 kbar. A mixture of Mg and B was put into a BN container of 6 mm internal diameter and 7 mm length. First, pressure was applied using a pyrophylite cube as a medium, then the temperature was increased during one hour, up to the maximum of 1700-2200°C, kept for 1-3 hours, and decreased during 1-2 hours. The result of a typical cubic anvil growth experiment is displayed in Fig. 2a), showing a collection of $MgB_2$ and BN crystals, sticking together. $MgB_2$ plate like golden crystals (see Fig. 2b) for example) were up to 1.5x0.9x0.2mm$^3$ in size and up to 230 µg in weight. In some experiments, hexagonal black crystals were obtained from the crystal growth process. Structural x-ray studies showed those to be a new nitride phase, namely $MgNB_9$ [Fig.2c)]. The structure of this compound consists of two kinds of boron polyhedra separated by Mg and N atoms [7,15].

**Thermodynamics of the Mg-B-N system at high pressure**

In order to determine the conditions of crystal growth, a basic knowledge of phase relationship in the Mg-B-N system is necessary. The pressure dependence of the melting temperatures of two eutectics and the decomposition temperature of $MgB_2$ were published by S. Lee *et al.* [16]. They concluded that $MgB_2$ decomposes at $T$>1200°C and 1600°C at $P$=20 kbar and 50 kbar, respectively. At $T$>1100°C, BN reacts with Mg, and therefore, at higher temperatures, the reactions in the ternary Mg-B-N system, shown in Fig. 3a), have to be considered. All our experiments have been performed with a composition with an excess of Mg over that in $MgB_2$. Typically, we used a ratio Mg:B equal to 1:1.2. The excess of Mg produces a high Mg vapor pressure. During the processes of crystal growth and synthesis of $MgB_2$ in a BN crucible, the $MgNB_9$ phase appears in some experiments. Figure 3b) shows the *P-T* phase diagram, with the phases obtained in our experiments for a composition with excess of Mg. Additionally, the BN hexagonal phase and Mg have been present in all samples (not shown). In experiments performed at *P*<15 kbar, only $MgNB_9$ crystals grew at *T*>1300°C. At *P*=15 kbar both $MgNB_9$ and $MgB_2$ crystals grew. After short processes, at pressure $P \geq 20$ kbar where the maximum temperature has been kept only for 10-15 min., we found $MgB_2$ crystals grown inside of $MgNB_9$ black crystals [Fig.2c)]. After longer times at high temperature and pressure $MgNB_9$ disappeared and only $MgB_2$ and BN crystals, sticking together, remained in the crucible [Fig.2a)]. The crystal growth process of $MgB_2$ is not a simple growth from a solution in molten magnesium. $MgB_2$ crystals are the products of a reaction in the ternary system Mg-B-N. On the first stage of the growth crystals grow by a peritectic decomposition of the intermediate nitride phase $MgNB_9$. The source of nitrogen is the BN, which reacts with molten Mg and B, forming $MgNB_9$. This compound decomposes at high temperature to $MgB_2$, according to the reaction:

(1)     $4Mg + 8B + BN \rightarrow MgNB_9 + 3Mg \rightarrow MgB_2 + BN + 6B + 3Mg \rightarrow 4MgB_2 + BN$

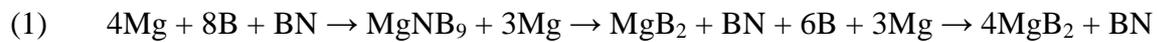

The $MgNB_9$ decomposition reaction takes places only at high pressure $P \geq 15$ kbar. At lower pressure, $MgNB_9$ remains as the final product. Pressure induces a decomposition of the $MgNB_9$ phase. The volume of the products of the decomposition reaction is 1% lower than the volume of the $MgNB_9$ compound, which can promote the decomposition at high pressure conditions. Products of the $MgNB_9$ decomposition are $MgB_2$ and BN crystals as well as boron. Boron reacts with magnesium and already grown $MgB_2$ crystals grow further until the whole boron is used. In the crucible after the growth process no boron can be found, only $MgB_2$, BN, Mg and in some of experiments in the cold part of the crucible crystals of $MgNB_9$.

The $MgB_2$ phase is unexpectedly stable up to very high temperatures, at least up to 2190°C. This is in disagreement with the data of S. Lee *et al.* [16], who reported the decomposition of $MgB_2$ at temperatures just above 1200°C, at *P*=25-30 kbar. We are growing $MgB_2$ crystals at these pressures even at temperatures as high as 2190°C. Most likely, the reason for this discrepancy is the fact that for the stability of the $MgB_2$ phase at T>1200°C, high hydrostatic pressure alone is not enough. According to Fig. 1b), high magnesium vapor pressure is necessary in addition to high hydrostatic pressure. Since we are working with an excess of Mg in a homogeneous precursor mixture, a high magnesium vapor pressure is produced, which stabilizes the $MgB_2$ phase. Small amounts of the $MgB_4$ phase, observed in the samples treated at high temperatures (above 1200$^o$C), disappears at T>1700$^o$C, and only $MgB_2$ remains as a stable phase.

**Magnetic properties of $MgB_2$ crystals**

Magnetization measurements performed with a SQUID magnetometer show, for the majority of the grown crystals, sharp superconducting transitions at 37-39K with widths of 0.3-0.5K. Nevertheless, in some cases we observe larger differences in $T_c$ values between various crystals, even up to several Kelvin. We identified several reasons for the lower $T_c$ of some crystals. Figure 4 a) shows the magnetization $M(T)$ curves for two crystals obtained with identical conditions ($P$=25 kbar, $T$=1900$^o$C, time $t$=2 h at the maximum temperature). The difference between two crystals grown in these processes originated from a difference of the boron purity: in one case, we have used crystalline boron 99.7%, in second amorphous boron 99.99% purity (both from Alfa Aesar). The resulting difference in $T_c$ is about 2K, caused by metallic impurities introduced into the crystals from boron with lower purity. Another case of lowering $T_c$ is shown on Figure 5 b). Here, two magnetization curves of $MgB_2$ crystals are displayed. The crystals were obtained in similar experiments performed at 30 kbar with higher (2190$^o$C) and lower (1850$^o$C) maximum growth temperatures. The magnetization curve of the crystal grown at 2190$^o$C shows a two step transition with a lower onset $T_c$. This is possibly due to the formation of defects in the structure of $MgB_2$ indicating, that the temperature of 2190$^o$C is close to the limit of stability of the $MgB_2$ phase at the conditions of the experiment.

**Structural investigations**

The structure of several crystals has been investigated with a x-ray (Nonius Kappa CCD) diffractometer, using graphite monochromated Ag Kα radiation [7]. The crystals had the form of platelets with dimensions between 0.12x0.12x0.03mm$^3$ and 0.4x0.4x0.06mm$^3$. The transition temperature of crystals from four different batches varied from 34.4K to 38.8K. In all cases, roughly 65 unique reflexions with $\sin\theta / \lambda > 0.2$ were used for structure refinements, and the obtained $R$ and $Rw$ agreement factors were between 1.5 and 2%, proving good quality of refinement of the structural parameters. The atoms were placed in their standard positions for the AlB$_2$-type structure (Mg (0 0 0), B(2/3 1/3 ½). Unexpectadly, the structural parameters obtained for the four samples were all equal within two estimated standard deviations. Their values are: $a$ = 3.085(1) Å, $c$ = 3.518(2)Å, anisotropic displacement parameters: $U_{11}$(Mg) = 0.0045(1)Å$^2$, $U_{33}$(Mg) = 0.0048(2) Å$^2$, $U_{11}$(B) = 0.0034(2) Å$^2$ and $U_{33}$(B) = 0.0044(2) Å$^2$. This agreement between structural data of the crystals from different batches indicates that the changes of $T_c$ observed between crystals grown in different conditions cannot be attributed to detectable structural modifications. Refining the B atom occupancy lead to values equal to 1.00(1). However, we note that a minor substitution of boron by carbon or nitrogen from the heater or crucible would be almost impossible to detect with x-ray diffraction, because of the small contrast between these elements, and therefore cannot be excluded here. Electron energy loss spectroscopy [17] and energy dispersive x-ray measurements did not show such impurities, therefore the differences in $T_c$ have to be atributed to not yet defined defects or impurities.

**Conclusions**

MgB$_2$ single crystals of millimeter size have been grown from a peritectic decomposition reaction of the magnesium bornitride MgNB$_9$. The MgB$_2$ crystals have been grown at temperatures up to 2200$^o$C. MgB$_2$ is stable up to the unexpectedly high temperature of 2190$^o$C, at high hydrostatic pressure, due to the high pressure of magnesium vapors produced by the excess of magnesium in the precursor. The MgNB$_9$ phase is not stable at high hydrostatic pressure at $P$>15 kbar at high temperature, but it appears in short processes at higher pressure as a metastable phase. After the short time of 10-20 minutes, it decomposes, and simultaneously MgB$_2$ and hexagonal BN crystals grow. $T_c$ of the crystals is sensitive to doping from impurities during crystal growth. Structural investigations proved a full occupation of the lattice positions and SQUID investigations showed sharp magnetic transitions, which demonstrates a high quality of the crystals.

**Acknowledgements**

This work was supported by the Swiss National Science Foundation, by the Swiss Federal Office BBW (contract Nr.02.0362), by the European Community (contract ICA1-CT-2000-70018) and by the Polish State Committee for Scientific Research (5 P03B 12421).

**Figure Captions**

**Fig. 1.** a) Temperature-composition phase diagrams of the Mg-B system under the Mg vapor pressure of 1 bar. b) Mg vapor pressure-temperature phase diagram for the Mg:B atomic ratio $x_{Mg}/x_B \geq 1/2$. The region of Liquid+$MgB_2$ represents the thermodynamic stability window for the growth of $MgB_2$ crystals from solution in Mg. All data are taken from Ref. [11].

**Fig. 2.** a) Sample containing $MgB_2$ and BN crystals after synthesis. b) Single crystals of $MgB_2$. c) Crystal of the new phase $MgNB_9$ with golden $MgB_2$ crystals grown inside. Scale size is 1mm.

**Fig. 3.** a) Ternary Mg-B-N system with compounds involved in the crystal growth of $MgB_2$ [7]. Field P corresponds to the composition of precursor used for crystal growth. b) *P-T* phase diagram of Mg-B-N system for a composition with excess Mg (Mg:B=1:1.2). Symbols show phases observed in the samples. Regions of appearance of various phases: A: $MgNB_9$, B: $MgB_2$, C: $MgB_2$+$MgB_4$. Additionally, BN and Mg were also present in all samples (not shown on diagram). $MgB_2$ single crystals have been grown in the region B. At pressures *P*>15 kbar the $MgNB_9$ compound appears only as a metastable phase.

**Fig. 4.** a) Magnetization curves of two $MgB_2$ crystals using the same growth conditions from boron of various purity. b) Magnetization curves of two $MgB_2$ crystals obtained in similar processes with various maximum temperatures.

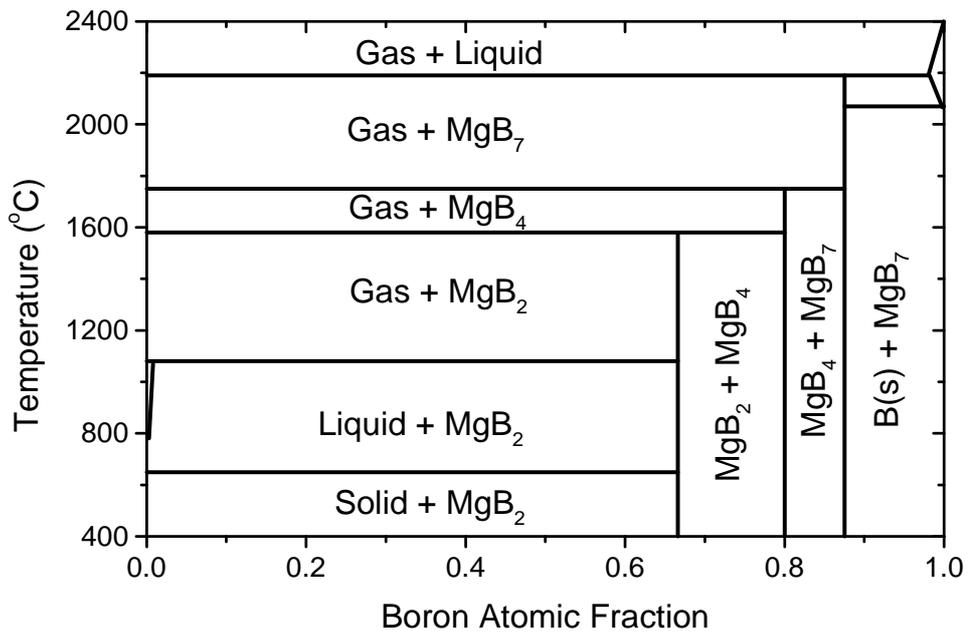

a)

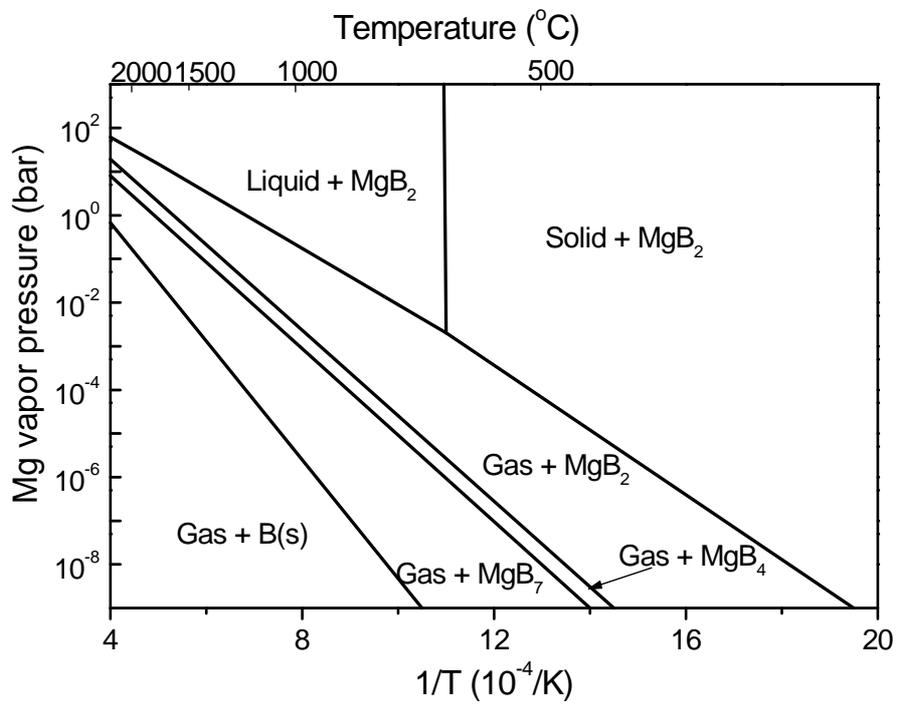

b)

Figure 1

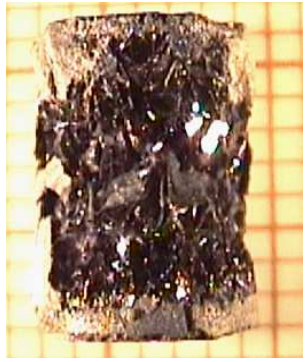

a)

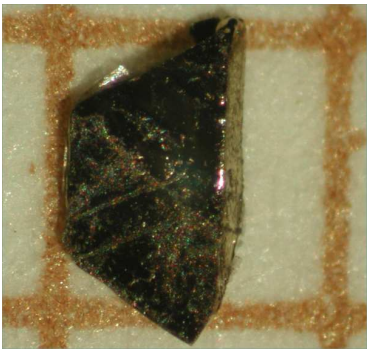 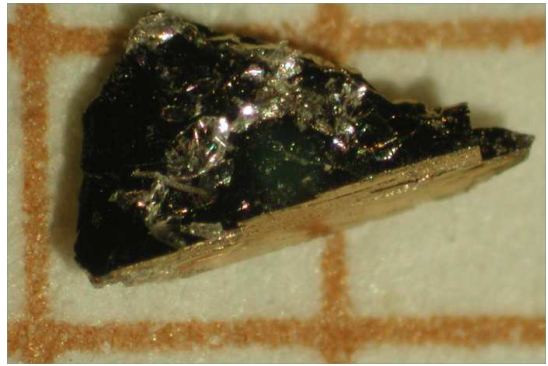

b)

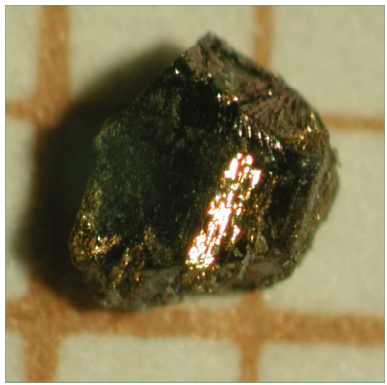

c)

Figure 2

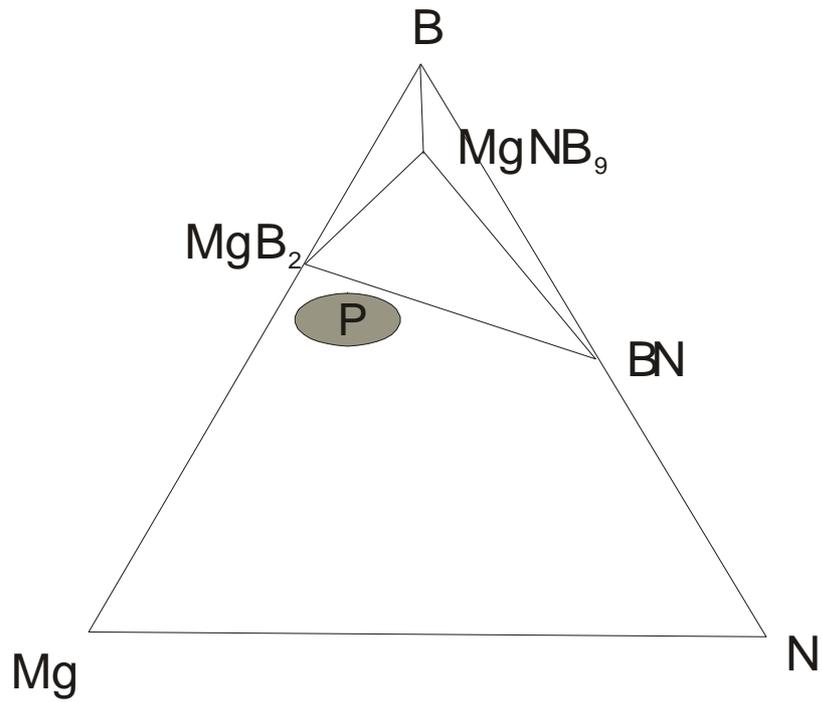

a)

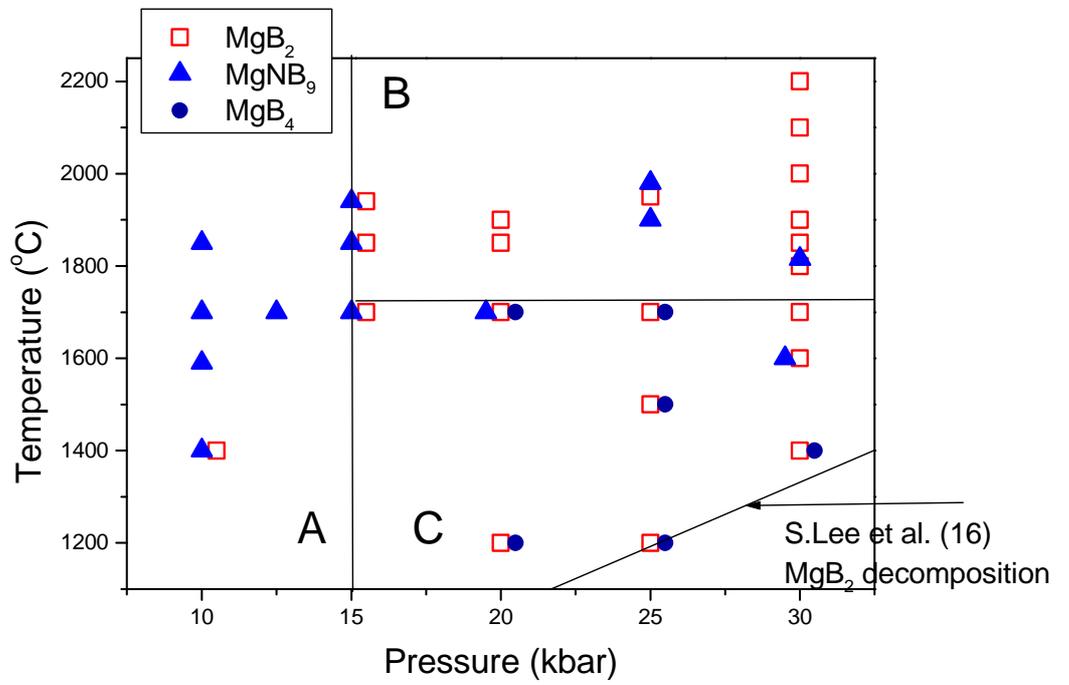

b)

Figure 3

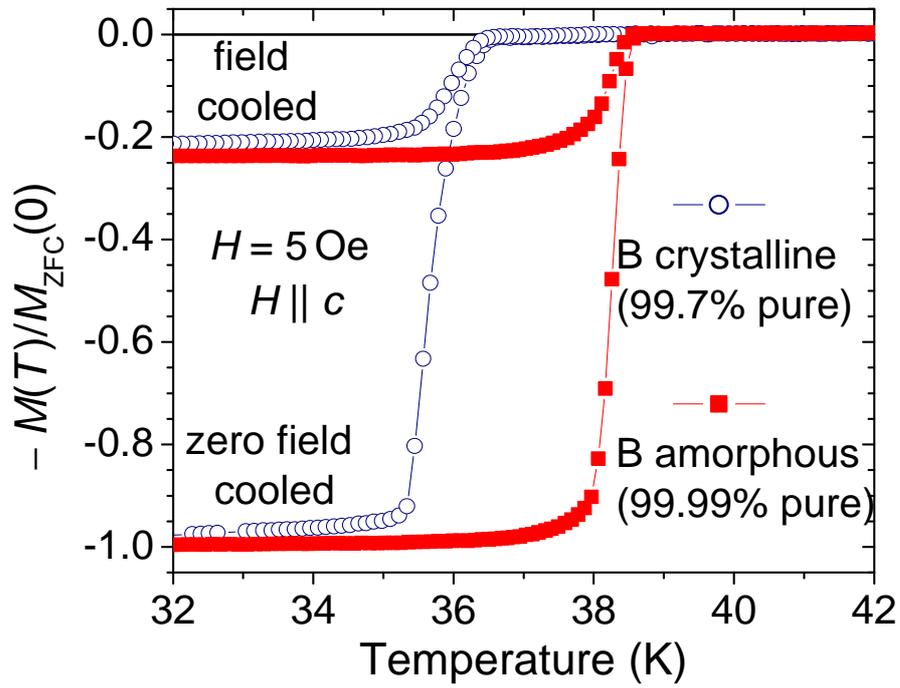

a)

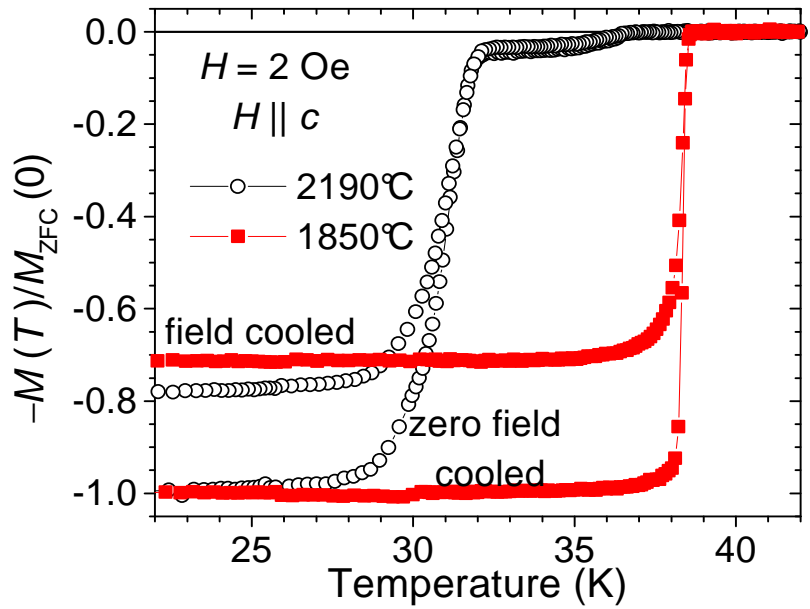

b)

Figure 4